\documentclass[aps,pra,
       reprint,
       superscriptaddress,
       showpacs,
   preprintnumbers,
   ]{revtex4-1}
\usepackage{amsfonts}
\usepackage{amssymb}
\usepackage{amsmath}
\usepackage{color}
\usepackage{graphicx}
\usepackage{subfigure}
\usepackage[pdftex,colorlinks=true,linkcolor=red,citecolor=blue]{hyperref}
\usepackage{color}
\usepackage{epsfig}
\usepackage{graphicx}
\usepackage{time}
\newcommand{\rd}{\mathrm{d}}               
\newcommand{\rD}{\mathrm{D}}               
\newcommand{\defby}{\equiv}                
\newcommand{\calN}{\mathcal{N}}            
\newcommand{\calL}{\mathcal{L}}            
\newcommand{\Expect}[1]
   {\ensuremath{\langle \, #1 \,  \rangle}}
\newcommand{\Comm}[2]
   {\ensuremath{[ \, #1, #2 \, ]}}
\newcommand{\AntiComm}[2]
   {\ensuremath{\{ \, #1, #2 \, \}}}
\newcommand{\Ln}[1]{\ln [ \, #1 \, ]}              
\begin{document}
%
%
%

\preprint{LA-UR-11-02248}

\title[BEC]
   {Analytical limits for cold atom Bose gases with tunable interactions}

\author{Bogdan Mihaila}
\affiliation{
   Los Alamos National Laboratory,
   Los Alamos, NM 87545}

\author{Fred Cooper}
\affiliation{
   Los Alamos National Laboratory,
   Los Alamos, NM 87545}
\affiliation{Santa Fe Institute,
   Santa Fe, NM 87501}

\author{John F. Dawson}
\affiliation{Department of Physics,
   University of New Hampshire,
   Durham, NH 03824}

\author{Chih-Chun Chien}
\affiliation{
   Los Alamos National Laboratory,
   Los Alamos, NM 87545}

\author{Eddy Timmermans}
\affiliation{
   Los Alamos National Laboratory,
   Los Alamos, NM 87545}


\begin{abstract}
We discuss the equilibrium properties of dilute Bose gases using a non-perturbative formalism based on auxiliary fields related to the normal and anomalous densities. We show analytically that for a dilute Bose gas of weakly-interacting particles at zero temperature, the leading-order auxiliary field (LOAF) approximation leads to well-known analytical results.  Close to the critical point the LOAF predictions are the same as those obtained using an effective field theory in the large-$N$ approximation.  We also report analytical approximations for the LOAF results in the unitarity limit, which compare favorably with our numerical results. LOAF predicts that the equation of state for the Bose gas in the unitarity limit is $E / (p V) = 1$, unlike the case of the Fermi gas when $E / (p V) = 3/2$.
\end{abstract}
\date{\today, \now}
\maketitle
%
%
\section{Introduction}

Recently we introduced  a new theoretical framework for the study of a dilute gas of Bose particles with tunable interactions\cite{PhysRevLett.105.240402} based on a loop expansion of the one-particle irreducible (1-PI) effective action in terms of composite-field propagators. The auxiliary field (AF) formalism makes use of the Hubbard-Stratonovitch transformation\cite{r:Hubbard:1959kx,*r:Stratonovich:1958vn} to rewrite the Lagrangian in terms of auxiliary fields related to the normal and anomalous densities. 
Employing general quantum field theoretical methods\cite{PhysRevD.10.2491,PhysRevD.10.3322,r:Bender:1977bh,r:Negele:1988fk,r:Calzetta:2008pb}, the AF formalism is part of a continuing effort in the community to apply methods traditionally used in high-energy physics\cite{PhysRevLett.71.3202,PhysRevB.55.15153,PhysRevB.56.14745,PhysRevA.69.033610,r:gasenzer:2005,r:gasenzer:2006,r:gasenzer:2007,PhysRevA.75.013613,PhysRevB.81.235108,PhysRevA.77.053603,PhysRevB.78.174528} to the study of ultracold atomic gases\cite{RevModPhys.76.599}. 

For an interacting dilute Bose gas, the leading-order auxiliary field (LOAF) approximation is a non-perturbative, conserving and gapless approximation that describes a large interval of values of the coupling constant, satisfies Goldstone's theorem and yields a second-order phase transition to a Bose-Einstein condensate (BEC) regime.  In contrast with other resummation schemes, such as the large-$N$ expansion\cite{PhysRevLett.83.1703,r:Baym:2000fk} or the functional renormalization techiques\cite{PhysRevA.77.053603,PhysRevB.78.174528,PhysRevB.81.235108}, here we treat the normal and anomalous densities on equal footing.  LOAF produces the same slope of the linear departure of the critical temperature from the noninteracting limit derived by Baym \emph{et al.}\cite{r:Baym:2000fk} using a large-N expansion for the critical theory. Unlike the large-N expansions developed by Baym \emph{et al.}, the LOAF approximation can be used at all temperatures. Furthermore, one can systematically improve upon the LOAF approximation  by  calculating the 1-PI action order-by-order corrections. The broken $U(1)$ symmetry Ward identities guarantee the preservation Goldstone's theorem order-by-order\cite{r:Bender:1977bh}. 

The detailed derivation of the LOAF approximation was discussed recently for the case of dilute Bose\cite{PhysRevA.Bose:2011} and Fermi\cite{PhysRevA.Fermi:2011} atomic gases. Unlike the case of Fermi gases where the LOAF approximation is equivalent to the standard Bardeen-Cooper-Schrieffer  (BCS) ansatz \cite{PhysRevLett.71.3202,PhysRevB.55.15153}, in the case of Bose gases the LOAF approximations leads to yet unexplored possibilities. Therefore it is important to study analytically the LOAF predictions for Bose gases in limiting cases such as the case of weakly-interacting systems and also in the unitarity limit, which corresponds to the strongly-interacting regime where the s-wave scattering length,  $a_0$, is much larger than the inter-particle distance. In the unitarity limit the properties of the system have a universal character\cite{PhysRevLett.99.090403}. The intrinsic non-perturbative character of the LOAF approximation may become particularly relevant because the development of novel cold atom technology that produce stable, flat potentials bound by a sharp edge\cite{r:Henderson:2006fv,r:Henderson:2009dz} leads to  the prospect of studying finite temperature properties of dilute gases, such as the BEC transition temperature, $T_{c}$, superfluid to normal fluid ratio, depletion, and specific heat, at fixed particle density~$\rho$. 

In this paper we focus on the study of the LOAF predictions in the broken-symmetry phase. We make contact with existing analytical approximations in the weakly-interacting limit, such as those discussed in the textbook of Fetter and Walecka\cite{r:Fetter:1971fk} and the analytical results obtained close to the critical temperature by Kita\cite{r:Kita:2005fk,r:Kita:2005uq,r:Kita:2006kx} using the related Luttinger-Ward functional. We will also show that the analytical techniques developed here can be used to study the LOAF predictions in the unitarity limit.

This paper is organized as follows: In Sec.~\ref{Seff.ss:AF}, we briefly review the derivation of the LOAF equations. The LOAF effective potential and the derivation of thermodynamic properties  are outlined in Sec.~\ref{Seff.ss:Veff}. In Sec.~\ref{bsp}  we specialize to the study of the interacting Bose properties in the broken-symmetry phase. In Sec.~\ref{Tzero} we study the zero-temperature LOAF results in the weakly-coupling limit and we compare with the weakly interacting Bose gas theory discussed by Fetter and Walecka\cite{r:Fetter:1971fk}. The zero-temperature analysis suggests the scaling of the LOAF equations discussed in Sec.~\ref{scaled}. In Sec.~\ref{critical} we discuss the LOAF properties close to the critical temperature in the weakly-interacting limit and we compare with the results obtained by Kita using a related approximation\cite{r:Kita:2006kx}. Analytical approximations of the LOAF predictions in the unitarity limit are discussed in Sec.~\ref{unitarity}. We conclude in Sec.~\ref{concl}.

%
%
\section{\label{Seff.ss:AF}Leading order auxiliary field (LOAF) formalism}

The detailed derivation of the LOAF approximation for the case of dilute Bose gases was discussed recently in Ref.~\onlinecite{PhysRevA.Bose:2011}. For completeness, we will review next the salient aspects of the AF-formalism derivation.

In dilute bosonic gas systems, the classical action is given by
\begin{equation}\label{BEC.aux.e:actionI}
   S[\, \phi,\phi^{\ast} \, ]
   =
   \int \! \rd x \> 
   \calL[ \, \phi,\phi^{\ast} \, ] 
   \>,
\end{equation}
with $\rd x \defby \rd t \, \rd^3 x$ and the Lagrangian density
\begin{gather}
   \calL[ \, \phi,\phi^{\ast} \, ]
   =
   \frac{i \hbar}{2} \, 
   [ \, 
      \phi^{\ast}(x) \, ( \partial_t \, \phi(x) )
      -
      ( \partial_t \, \phi^{\ast}(x) ) \, \phi(x) \,
   ]
   \notag \\
   {}-
   \phi^{\ast}(x) \, 
   \Bigl ( \,
      -
      \frac{\hbar^2\nabla^2}{2m}
      -
      \mu \,
   \Bigr ) \, 
   \phi(x)
   -
   \frac{\lambda_0}{2} \, | \, \phi(x) |^4 \>.
   \label{BEC.aux.e:LagI}
\end{gather}
Here, $\mu$ is the chemical potential and $\lambda_0$ is the bare coupling constant.
In the auxiliary field formalism we use the Hubbard-Stratonovitch transformation\cite{r:Hubbard:1959kx,r:Stratonovich:1958vn} to eliminate the quartic interaction in Eq.~\eqref{BEC.aux.e:LagI} by introducing the real and complex auxiliary fields (AF),  $\chi(x)$ and $A(x)$, related to the normal and anomalous densities.
We add to Eq.~\eqref{BEC.aux.e:LagI} the AF Lagrangian density\cite{PhysRevD.10.2491,PhysRevD.10.3322,r:Bender:1977bh}
\begin{align}
   \calL_{\text{aux}}[\phi,\phi^{\ast},\chi,A,A^{\ast}]
   = &
   \frac{1}{2 \lambda_0} \,
   \Bigl ( \,
      \chi(x) - \sqrt{2} \, \lambda_0 \, | \phi(x) |^2 \,
   \Bigr )^2
   \notag \\
   & -
   \frac{1}{2 \lambda_0} \,
   \Bigl | \,
      A(x) 
      - 
      \lambda_0 \, \phi^{2}(x) \,
   \Bigr |^2
   \>.
   \label{BEC.aux.e:Laux}
\end{align}
Then, the action becomes
\begin{align}
   &S[\Phi,J] =S[ \phi_a,\chi,A,A^\ast, j_a,s,S]
   \label{BEC.aux.e:actionII} \\
   & \quad 
   =
   - \frac{1}{2} \, 
   \iint \rd x \, \rd x' \,
   \phi_a(x) \, G^{-1}{}^a{}_b[\chi,A](x,x') \, \phi^b(x')
   \notag \\
   & \qquad {}+
   \int \rd x \,
   \bigl \{ \,
      \bigl [ \,
         \chi^2(x) - | A(x) |^2 \,
      \bigr ] / (2\lambda_0)
      -
      s(x) \chi(x)
      \notag \\
      &
      +
      S^{\ast}(x) A(x)
      +
      S(x) A^{\ast}(x)
      +
      j^{\ast}(x) \phi(x)
      +
      j(x) \phi^{\ast}(x) \,
   \bigr \} \>,
   \notag
\end{align}
with
\begin{align}
   &G^{-1}{}^a{}_b[\chi,A]
   \label{BEC.aux.e:G0invdef} 
   \\ \notag & 
   = 
   \delta(x,x') \! 
   \begin{pmatrix}
      -
      \gamma \,  \nabla^2
      \! -
      i \hbar \, \partial_t
      + \chi'
      & - A(x) \\
      - A^{\ast}(x) & 
      -
      \gamma \, \nabla^2
      \, +
      i \hbar \, \partial_t
      + \chi'
   \end{pmatrix} \! ,
\end{align}
where we introduced the notations $\gamma = \hbar^2 / (2m)$ and
\begin{equation}
   \chi' =  \sqrt{2} \, \chi(x)  -  \mu 
   \>,
\end{equation}
together with a two-component notation, $\phi^a(x) = \{ \, \phi(x), \phi^{\ast}(x) \, \}$, for $a = 1,2$.  
$\Phi(x)$ and $J(x)$ signify the five-component fields and currents.  

The generating functional for connected graphs is
\begin{equation}\label{BEC.Seff.e:Z}
   Z[J]
   =
   e^{i W[J] / \hbar}
   = 
   \calN
   \int \rD \Phi \>
   e^{ i S[\Phi;J] / \hbar } \>,
\end{equation}
with $S[\Phi;J]$ given by Eq.~\eqref{BEC.aux.e:actionII}.  
Performing the path integral over the fields $\phi_a$, we obtain the effective action for $\chi, A,A^\ast$, as
\begin{align}
   & S_{\text{eff}}[ \chi;J,\epsilon ]
   =
   \frac{1}{2 } \iint \rd x \, \rd x' \,
   j_{a}(x) \, G[\chi]^a{}_b(x,x') \, j^a(x)
   \\ \notag
   &{}+
   \int \rd x \,
   \Bigl \{ \,
      \frac{\chi_i(x) \, \chi^{i}(x)}{2\lambda_0}
      -
      S_{i}(x) \, \chi^{i}(x)
      - \frac{\hbar}{2i}  
      \text{Tr} \, \Ln{ G^{-1} } \, 
   \Bigr \} \>,
\end{align}
where 
\begin{align}
   \chi^{i}(x)
   &
   =
   \bigl \{ 
      \chi(x), A(x)/\sqrt{2}, A^{\ast}(x)/\sqrt{2} 
   \bigr \} 
   \>,
   \\
   S^{i}(x)
   &
   =
   \bigl \{ 
      s(x), S(x)/\sqrt{2}, S^{\ast}(x)/\sqrt{2} 
   \bigr \} 
   \>.
\end{align}

Next, we expand the effective action about the stationary points, $\chi_0^{i}(x)$, defined by $\delta S_{\text{eff}}[ \chi;j ] /  \delta \chi_{i}(x) = 0$.
We obtain the ``gap'' equations:
\begin{align}
   \frac{\chi_0(x)}{\lambda_0}
   &=
   \sqrt 2 \,
   \Bigl [
      | \phi_0(x) |^2
      +
      \frac{\hbar}{2i} \, \mathrm{Tr} \, G(x,x)  
   \Bigr ]
   + s(x) \>,
  \\
  \frac{A_0(x)}{\lambda_0}
   &=
   \Bigl [
       \phi^2_0(x) 
      +
      \frac{\hbar}{i} \, G^{2}{}_{1}(x,x) 
   \Bigr ] 
   + S(x) \>,
\end{align}
where we introduced the notations $\phi^a_0[\chi_0](x)$ as
\begin{equation}\label{BEC.Seff.e:phi0def}
   \phi^a_0[\chi_0](x)
   =
   \int \rd x' \, G[\chi_0]^a{}_b(x,x') \, j^b(x') \>. 
\end{equation}
Both $\chi_0$ and $A_0$ include self-consistent fluctuations. 

Expanding the effective action about the stationary point, we write 
\begin{align}
   &S_{\text{eff}}[ \chi;J ]
   =
   S_{\text{eff}}[ \chi_0;J ]
   +
   \frac{1}{2} \iint \rd^4 x \, \rd^4 x' \,
   D_{ij}^{-1}[\chi_0](x,x')
   \notag \\ & \qquad {} \times
   \label{BEC.Seff.e:Seffexpand} 
   [ \chi^i(x) - \chi^i_0(x) ] \,
   [ \chi^j(x') - \chi^j_0(x') ]
   +
   \dotsb
   \>,
\end{align}
where $D_{ij}^{-1}(x,x')$ is given by the second-order derivatives,
\begin{equation}
   D_{ij}^{-1}[\chi_0](x,x')
   =
   \frac{ \delta^2 \, S_{\text{eff}}[ \chi^a] }
        { \delta \chi^i(x) \, \delta \chi^j(x') } \, \bigg |_{\chi_0} 
   \>,
\end{equation}
evaluated at the stationary points.  
By keeping the gaussian fluctuations and Legendre transforming, the one-particle irreducible (1-PI) graphs generating functional
\begin{align}
   &\Gamma[\Phi]
   =
   \int \rd x \, j_{\alpha}(x) \, \phi^{\alpha}(x)
   -
   W[J]
   \label{BEC.Seff.e:vertexfctdef} \\
   &
   =
   \frac{1}{2} \iint \rd x \, \rd x' \,
   \phi_a(x) \, G^{-1}[\chi]^{a}{}_{b}(x,x') \, \phi^b(x')
   \notag \\
   &\quad {}-
   \int \rd x \,
   \Bigl \{ \,
      \frac{\chi_{i}(x) \, \chi^{i}(x)}{2\lambda_0}
      - \frac{\hbar}{2i}  
      \text{Tr} \bigl \{ \,
         \Ln{ G^{-1}[\chi](x,x) } \,
      \bigr \} 
      \notag \\
      &\qquad \qquad \qquad {}-
      \frac{\hbar}{2i} \,
      \text{Tr} \, \Ln{ D_{ii}^{-1}[\Phi](x,x) } \, 
   \Bigr \}
   +
   \dotsb
   \>,
   \notag    
\end{align}
is the negative of the classical action plus self-consistent one-loop corrections in the $\phi_a$ and $\chi_i$ propagators. 
The last term in Eq.~\eqref{BEC.Seff.e:vertexfctdef} is next-to-leading order \cite{r:Bender:1977bh}, and is not included in the leading-order auxiliary field (LOAF) approximation. Hence, the static part of the effective action per unit volume is 
\begin{align}
   V_{\text{eff}}[\Phi]
   &=
   \chi'  \, | \phi |^2
   -
   \frac{1}{2} \,
   (
      A^{\ast} \, \phi^2 
      +
      A \, \phi^{\ast\,2} 
   ) 
   \notag \\
   & \qquad{}-
      \frac{\chi^2 - | A |^2}{2\lambda_0}
      + 
      \frac{\hbar}{2i}  
      \text{Tr} \, \Ln{ G^{-1}[\chi ] } \, 
      \>.
   \label{BEC.Seff.e:Veff}    
\end{align}
In the imaginary time formalism, the last term in Eq.~\eqref{BEC.Seff.e:Veff}  becomes
\begin{equation}
   \frac{\hbar}{2i}  
   \text{Tr} \, \Ln{ G^{-1}[\chi ] } 
   = \!\!
   \int \frac{\rd^3 k}{(2\pi)^3} \,
   \Bigl \{ \,
      \frac{\omega_k}{2}
      +
      \frac{1}{\beta} \,  \Ln{ 1 - e^{-\beta \omega_k} } \,
   \Bigr \} \>,
\end{equation}
where the dispersion relation is given by
\begin{equation}
    \omega_k^2 = ( \epsilon_k + \chi' )^2 -  |A|^2 
    \>,
\end{equation} 
with $\epsilon_k = \gamma k^2$. At the minimum of the effective potential, we also have
\begin{equation}\label{BEC.Seff.e:brokencase}
   \frac{\delta V_{\text{eff}}[\Phi]}{\delta \phi^{\ast}} \Bigl |_{\phi_0}
   =
   \chi' \, \phi_0
   -
   A  \, \phi_0^{\ast}
   =
   0 \>.
\end{equation}
Using the $U(1)$ gauge symmetry, we choose $\phi_0$ to be real in the broken-symmetry phase. Then, $A$ is real and the dispersion, $\omega_k^2 = \epsilon_k ( \epsilon_k + 2 A )$, represents the Goldstone theorem.  

%
%
\section{\label{Seff.ss:Veff}Thermodynamics in the LOAF approximation}

Using standard regularization techniques for effective-field theories\cite{PhysRevC.59.2052,PhysRevLett.88.042504}, the renormalized effective potential 
\begin{align}
   &V_{\text{eff}}
   =
  (  \chi' - A )  \rho_0
   -
   \frac{ ( \chi' + \mu)^2}{4\lambda}
   +
   \frac{A^2 }{2\lambda}
\label{eq:veff}
   \\ \notag
   &
   +  \!\!
   \int \!\!
   \frac{d^3 k}{(2\pi)^3}
   \Bigl [
      \frac{1}{2}
      \Bigl (
         \omega_k
         -
         \epsilon_k
         -
         \chi'
         +
         \frac{A^2}{2 \epsilon_k}
      \Bigr )
      \! +
     T  \ln( 1 - e^{-\omega_k/T} )
   \Bigr ] ,
\end{align}
represents the grand potential per unit volume,
\begin{equation}
   V_{\text{eff}}  = \Omega[T,\mu,\mathcal{V}] / \mathcal{V}
   \>.
\end{equation}
Here, $\rho_0 = \phi_0^2$ is the condensate density, and the renormalized coupling constant is related to the s-wave scattering length by $\lambda = 8\pi \gamma \, a_0$.
The values of $\chi'$ and $A$ are obtained by solving self-consistently  the  gap equations
\begin{align}
\label{eq_1}
   \rho
   &=
   \rho_0
   + \!
   \int \! \frac{d^3 k}{(2\pi)^3}
   \Bigl (
      \frac{\epsilon_k + \chi'}{2\omega_k}
      -
      \frac{1}{2}
   \Bigr )
   + \!
   \int \! \frac{d^3 k}{(2\pi)^3}
      \frac{\epsilon_k + \chi'}{\omega_k} \,
      n_{\omega}
   \>,
   \\
\label{eq_2}
   \rho_0
   &=
   \frac{A}{\lambda}
   - \!
   \int \! \frac{d^3 k}{(2\pi)^3}
   \Bigl (
      \frac{A}{2\omega_k}
      -
      \frac{A}{2\epsilon_k}
   \Bigr )
   - \!
   \int \! \frac{d^3 k}{(2\pi)^3}
      \frac{A}{\omega_k} \, n_{\omega}
   \>,
\end{align}
where $n_\omega=1/ (e^{\omega_k/T} - 1)$ is the Bose-Einstein particle distribution and we assume $k_B=1$ units.

Using the grand potential, $\Omega[T,\mu,\mathcal{V}]$, we calculate
the total number of particles
\begin{equation}
N[T,\mu,\mathcal{V}]
   =
   - \partial_\mu \Omega |_{T,\mathcal{V}}
  \>,
\end{equation}
the pressure
\begin{equation}
\label{p_eq}
   p[T,\mu]
   =
   - \partial_\mathcal{V} \Omega |_{T,\mu}
  \>,
\end{equation}
entropy
\begin{equation}
   S[T,\mu,\mathcal{V}]
   =
   - \partial_T \Omega |_{\mu,\mathcal{V}}
  \>.
\end{equation}
The energy is obtained as 
\begin{equation}
   E = \Omega + T S + \mu N
   \>.
\end{equation}
Hence, the physical density is given by
\begin{equation}
   \rho
   =
   - \partial_\mu V_{\text{eff}}
   = \frac{1}{2\lambda} \ ( \chi' + \mu )
  \>,
\end{equation}
the pressure is
\begin{equation}
   p
   = - V_{\text{eff}}[\rho_0,\mu,T]
   \>,
\end{equation}
and the entropy density, $s = S/\mathcal{V}$, is
\begin{align}
   s
   & =
   \int
   \frac{d^3 k}{(2\pi)^3} \,
      \bigl [
      (1 + n_\omega ) \ln( 1 + n_\omega )
      -
      n_\omega \ln n_\omega
      \bigr ]
   \>,
   \notag \\ & =
   \int
   \frac{d^3 k}{(2\pi)^3} \,
      \Bigl [
      -
      \ln( 1 - e^{-\omega_k/T} )
      +
      \frac{\omega_k/T}{e^{\omega_k/T} - 1}
      \Bigr ]
   \>.
\end{align}
The energy density, $\varepsilon = E/\mathcal{V}$, is obtained as
\begin{equation}
    \varepsilon = - p + T s + \mu \rho
    \>.
\end{equation}

%
%
\section{\label{bsp}Broken symmetry phase}

In the following we will focus on the broken symmetry region of the phase diagram, i.e. the regime where the density of the BEC condensate is nonzero, $\rho_0 \neq 0$.
In the broken-symmetry phase, $T<T_c$, we have equal normal and anomalous densities, $\chi' = A = \Delta$, and the  dispersion relation becomes 
\begin{equation}\label{wk}
    \omega_k^2 = \epsilon_k ( \epsilon_k + 2 \Delta)
    \>.
\end{equation}
We note that in the long-wavelength limit, Eq.~\eqref{wk} reduces to the linear dispersion relation
\begin{equation}
    \omega_k \approx \hbar k \, \sqrt{ \frac{\Delta}{m} }
    \>,
    \quad k \rightarrow 0
    \>,
\end{equation} 
with the characteristic velocity (zero sound), 
$
    \sqrt{ \Delta/m }
$. 
Comparing~\eqref{wk} with Eq.~(21.11) in Ref.~\onlinecite{r:Fetter:1971fk}, i.e.
\begin{equation}
    \omega_k \approx \hbar k \, \sqrt{ \frac{\rho_0 \, V(0)}{m} }
    \>,
    \quad k \rightarrow 0
    \>,
\end{equation} 
we find that the parameter $\chi' = A =\Delta$ in LOAF plays the role of $\rho V(0)$ in the weakly interacting Bose gas theory discussed by Fetter and Walecka, with the zeroth moment of the potential
\begin{equation}
    V(0) = 4\pi \int V(r) \, r^2 \, dr \ \leftrightarrow \ \frac{\Delta}{\rho_0}
    \>.
\end{equation}

In the broken-symmetry phase we can calculate explicitly the temperature independent integrals in Eqs.~\eqref{eq:veff}, \eqref{eq_1} and~\eqref{eq_2}, as
\begin{align}
   \int \! \frac{d^3 k}{(2\pi)^3}
   \Bigl (
      \frac{\epsilon_k + \Delta}{2\omega_k}
      -
      \frac{1}{2}
   \Bigr )
   & =
   \frac{\sqrt 2}{3} \, \mathcal{A} \, \Delta^{3/2}
   \>,
   \\
   \int \! \frac{d^3 k}{(2\pi)^3}
   \Bigl (
      \frac{\Delta}{2\omega_k}
      -
      \frac{\Delta}{2\epsilon_k}
   \Bigr )
   & =
   - \, \sqrt 2 \, \mathcal{A} \, \Delta^{3/2}
   \>,
   \\
   \int \! \frac{d^3 k}{(2\pi)^3}
      \frac{1}{2}
      \Bigl (
         \omega_k
         -
         \epsilon_k
         -
         \chi'
         +
         \frac{A^2}{2 \epsilon_k}
      \Bigr )
   & =
   \frac{8 \sqrt 2}{15} \, \mathcal{A} \, \Delta^{5/2}
   \>,
   \\
   \int \! \frac{d^3 k}{(2\pi)^3}
      \frac{1}{2}
      \Bigl (
         \frac{\epsilon_k}{\omega_k}
         + \frac{\Delta}{\epsilon_k}
         - 1
      \Bigr )
   & =
   \frac{4 \sqrt 2}{3} \, \mathcal{A} \, \Delta^{3/2}
   \>.
\end{align}
Here we introduced the notation $\mathcal{A} = (4\pi^2)^{-1} \gamma^{-3/2}$.
We have also
\begin{align}
   &
   \int \! \frac{d^3 k}{(2\pi)^3}
      \frac{\epsilon_k + \Delta}{\omega_k} \,
      n_{\omega}
   =
   \mathcal{A} T^{3/2} \!\!
   \int_0^\infty \!
      \frac{\epsilon + \delta}{\omega(\delta)} \,
      \frac{\sqrt \epsilon \, d \epsilon}{e^{\omega(\delta)}- 1}
   \>,
   \\
   &
   \int \! \frac{d^3 k}{(2\pi)^3}
      \frac{\Delta}{\omega_k} \, n_{\omega}
   =
   \mathcal{A} T^{3/2} \!\!
   \int_0^\infty \!\!
      \frac{\delta}{\omega(\delta)} \,
      \frac{\sqrt \epsilon \, d \epsilon}{e^{\omega(\delta)} - 1}
   \>,
   \\
   &
   \int \!\! \frac{d^3 k}{(2\pi)^3} \!
      \ln( 1 - e^{- \omega_k / T} )
   \! = \!
   \mathcal{A} T^{3/2} \!\!\!
    \int_0^\infty \!\!\!\!\!
      \sqrt \epsilon \, d \epsilon
      \ln \! \bigl [ 1 - e^{- \omega(\delta)} \bigr ]
      ,
\end{align}
with $\delta = \Delta/T$.
Hence, the gap equations~\eqref{eq_1} and~\eqref{eq_2} become
\begin{align}
   \rho
   & = \rho_0
   + \, \frac{\sqrt 2}{3} \, \mathcal{A} \, \Delta^{3/2}
   + \mathcal{A} \, T^{3/2} \!\!
   \int_0^\infty \!
      \frac{\epsilon + \delta}{\omega(\delta)} \,
      \frac{\sqrt \epsilon \, d \epsilon}{e^{\omega(\delta)}- 1}
   \>,
\label{eq:gap_bs1}
   \\
\label{eq:gap_bs2}
   \rho_0
   & =
   \frac{\Delta}{\lambda}
   + \, \sqrt 2 \, \mathcal{A} \, \Delta^{3/2}
   - \mathcal{A} \, T^{3/2} \!\!
   \int_0^\infty \!\!
      \frac{\delta}{\omega(\delta)} \,
      \frac{\sqrt \epsilon \, d \epsilon}{e^{\omega(\delta)} - 1}
   \>,
\end{align}
and the pressure~\eqref{p_eq} reads
\begin{align}
   p
   = &
   \lambda \rho^2
   -
   \frac{1}{2\lambda} \, \Delta^2
   -
   \frac{8 \sqrt 2}{15} \, \mathcal{A} \, \Delta^{5/2}
\label{eq:gap_p}
   \\ \notag &
   +
   \mathcal{A} \, T^{5/2}
    \int_0^\infty
      \sqrt \epsilon \, d \epsilon
      \ln \bigl [ 1 - e^{- \omega(\delta)} \bigr ]
   \>.
\end{align}
The entropy density is given by
\begin{align}
   s
\label{eq:gap_s}
   & =
   \mathcal{A} T^{3/2} \!\!
    \int_0^\infty \!\!\!\!\!
      \sqrt \epsilon \, d \epsilon \, 
      \Bigl \{
      -
      \ln \! \bigl [ 1 - e^{- \omega(\delta)} \bigr ]
      +
      \frac{\omega(\delta)}{e^{\omega(\delta)} - 1}
      \Bigr \}
   .
\end{align}

%
%
\section{\label{Tzero}Zero temperature properties in the weakly-interacting limit}

With the above results we can study the properties of the zero-temperature Bose gas. For $T=0$, the gap equations combine to give
\begin{align}
   \frac{\Delta_0}{\lambda \, \rho}
   & =
   1 - \frac{4 \sqrt 2}{3} \frac{\mathcal{A}}{\rho} \, \Delta_0^{3/2}
   \>,
\label{eq:Delta_0}
   \\
   \frac{\rho_0}{\rho}
   & =
   1 - \frac{\sqrt 2}{3} \, \frac{\mathcal{A}}{\rho} \, \Delta_0^{3/2}
   \>,
\label{eq:frac_0}
   \\
   p_0
   & =
\label{eq:p_0}
   \lambda \rho^2 - \frac{1}{2\lambda} \Delta_0^2
   - \frac{8 \sqrt 2}{15} \, A \, \Delta_0^{5/2}
   \>.
\end{align}
We also have
$
   \varepsilon_0
   =
   - p_0 + \mu_0 \rho
$,
with
$
   \mu_0
   =
   2 \lambda \rho - \Delta_0
$.
Here we note: $\lambda \rho = (8\pi \, \xi) \, \gamma \, \rho^{2/3}$, $\lambda \mathcal{A} = (2/\pi) \gamma^{-1/2} a_0$, with the dimensionless parameter, $\xi = \rho^{1/3}a_0$.

It is convenient to introduce the following rescaled variables
\begin{equation}
    \tilde \Delta = \frac{\Delta}{\lambda \rho} \>,
    \quad
    \tilde \mu = \frac{\mu}{\lambda \rho} \>,
    \quad
    \tilde p = \frac{p}{\lambda \rho} \>,
    \quad
    \tilde \varepsilon = \frac{\varepsilon}{\lambda \rho} \>.
\label{eq:T0_scaled}
\end{equation}
Then, we obtain
\begin{align}
   \tilde \Delta_0
   & =
   1 - \frac{32}{3 \sqrt \pi} \, \xi^{3/2} \, \tilde \Delta_0^{3/2}
   \>,
\label{eq:Delta_r0}
   \\
   \rho_0 / \rho
   & =
   1 - \frac{8}{3 \sqrt \pi} \, \xi^{3/2} \, \tilde \Delta_0^{3/2}
\label{eq:frac_r0}
   \>,
   \\
   \tilde p_0 / \rho
   & =
   1 - \frac{1}{2} \tilde \Delta_0^2
   - \frac{64}{15 \sqrt \pi} \, \xi^{3/2} \, \tilde \Delta_0^{5/2}
\label{eq:p_r0}
   \>,
\end{align}
and
\begin{equation}
   \tilde \varepsilon_0
   =
   - \tilde p_0 + \tilde \mu_0 \rho
   \>,
\label{eq:E_r0}
\end{equation}
with
\begin{equation}
   \tilde \mu_0
   =
   2 - \tilde \Delta_0
   \>.
\label{eq:mu_r0}
\end{equation}

%
\begin{figure}[t]
   \centering
   \includegraphics[width=0.9\columnwidth]{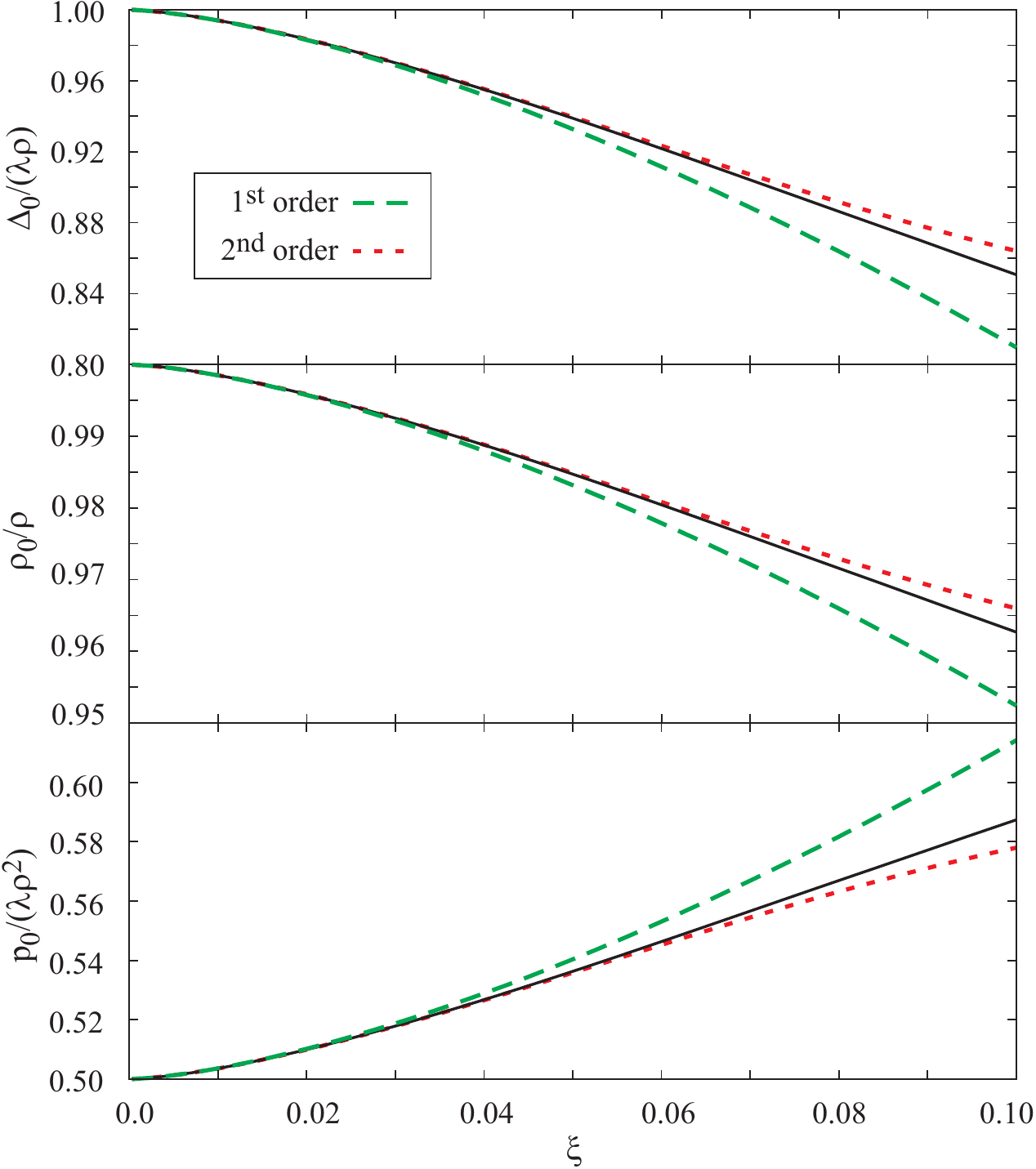}
   \caption{\label{Fig_T0_weak} (Color online) Comparisons of the exact zero-temperature values of the auxiliary field, $\Delta_0$, condensate fraction, $\rho_0/\rho$, pressure, $p_0$, and their respective first- and second-order approximations,  as a function of $\xi = \rho^{1/3}a_0$, in the weakly-interacting regime.
   The LOAF approximations for the auxiliary field, $\Delta_0$, condensate fraction, $\rho_0/\rho$, and pressure, $p_0$, in the weakly-interacting regime, are given in Eqs.~\eqref{T0:Delta0}, \eqref{T0:rho0} and \eqref{T0:p0}, respectively.
   }
\end{figure}
%

Iterating Eq.~\eqref{eq:Delta_r0}, we obtain the following successive approximations for $\Delta_0$,
\begin{align}
   \tilde \Delta_0
   & \approx
   1 - y
\label{eq:delta_app}
   \>,
   \\ \notag
   & \approx
   1 - y \, (1- y)^{3/2}
   \>,
   \\ \notag
   & \approx
   1 - y \, \bigl [1- y \, (1- y)^{3/2} \bigr ]^{3/2}
   \>, \cdots
\end{align}
with $y = 32/(3 \sqrt \pi) \, \xi^{3/2}$. Hence, we obtain
\begin{equation}
   \tilde \Delta_0
   =
   1 - \frac{32}{3\sqrt{\pi}} \, \xi^{3/2}
   + \frac{512}{3\pi} \, \xi^3
   + \cdots
   \>,
   \label{T0:Delta0}
\end{equation}
and
\begin{equation}
   \tilde \mu_0
   =
   1 + \frac{32}{3\sqrt{\pi}} \, \xi^{3/2}
   - \frac{512}{3\pi} \, \xi^3
   + \cdots
   \>.
\end{equation}
The coefficient of the $\xi^{3/2}$ term above is the same as in Eq.~(22.20) in Ref.~\onlinecite{r:Fetter:1971fk}.
Substituting Eqs.~\eqref{eq:delta_app} in Eq.~\eqref{eq:frac_r0}, we obtain
\begin{align}
   \rho_0 / \rho
   & \approx
   1 - \frac{y}{4} (1-y)^{3/2}
\label{eq:frac_0_app}
   \>,
   \\ \notag
   & \approx
   1 - \frac{y}{4} \, \bigl [ 1 - y \, (1-y)^{3/2} \bigr ]^{3/2}
   \>,
   \\ \notag
   & \approx
   1 - \frac{y}{4} \, \Bigl \{1 - y \, \bigl [ 1 - y \, (1-y)^{3/2} \bigr ]^{3/2} \Bigr \}^{3/2}
   , \cdots
\end{align}
which gives
\begin{equation}
   \rho_0 / \rho
   =
   1 - \frac{8}{3\sqrt{\pi}} \, \xi^{3/2}
   + \frac{128}{3\pi} \, \xi^3
   + \cdots
   \label{T0:rho0}
   \>.
\end{equation}
The coefficient of the $\xi^{3/2}$ term above is the same as in Eq.~(22.14) in Ref.~\onlinecite{r:Fetter:1971fk}.
Similarly, from Eqs.~\eqref{eq:p_r0} and~\eqref{eq:delta_app}, we obtain
\begin{equation}
   \tilde p_0 / \rho
   =
   1 -
   \frac{1}{2} \,
   \Bigl [
   (1 - y)^2
   + \frac{4}{5} \, y \, (1 - y)^{5/2}
   \Bigr ]
   \>,
   \cdots
\end{equation}
which gives
\begin{equation}
   \tilde p_0 / \rho
   =
   \frac{1}{2} + \frac{32}{5\sqrt{\pi}} \, \xi^{3/2}
   - \frac{1024}{9\pi} \, \xi^3
   + \cdots
   \>,
   \label{T0:p0}
\end{equation}
and
\begin{equation}
   \tilde \varepsilon_0 / \rho
   =
   \frac{1}{2} + \frac{64}{15\sqrt{\pi}} \, \xi^{3/2}
   - \frac{512}{9\pi} \, \xi^3
   + \cdots
   \>.
\end{equation}
Again, the coefficient of the $\xi^{3/2}$ term above is the same as in Eq.~(22.19) in Ref.~\onlinecite{r:Fetter:1971fk}. 

\textcolor{black}{We note that the next-to-leading order correction to the zero-temperature energy was calculated by Wu~\cite{PhysRev.115.1390}, yielding a second-order logarithmic term that cannot be captured by LOAF, which is only a one-loop approximation. The second-order logarithmic correction was later confirmed by Hugenhltz and Pines~\cite{PhysRev.116.489} and by Sawada~\cite{PhysRev.116.1344}.}

For illustrative purposes, in Fig.~\ref{Fig_T0_weak} we depict the exact zero-temperature values of the auxiliary field, $\Delta_0$, condensate fraction, $\rho_0/\rho$, pressure, $p_0$, and their respective first- and second-order approximations,  as a function of $\xi = \rho^{1/3}a_0$, in the weakly-interacting regime.

\textcolor{black}{By construction~\cite{PhysRevA.Bose:2011}, in the weak coupling limit  the LOAF approximation agrees with the Bogoliubov approximation\cite{r:Bogoliubov:1947ys,r:Andersen:2004uq}, which represents the leading-order low-density approximation of the theory. In addition, the related Popov approximation can be obtained from Eqs.~\eqref{eq_1} and~\eqref{eq_2} by setting $A =\chi' = \lambda \rho_0$ and neglecting the quantum fluctuations in the anomalous density. We showed in Ref.~\onlinecite{PhysRevA.Bose:2011} that the LOAF and the ``gapless" Popov approximation results become qualitatively similar in the weak coupling limit, even though the order of the phase transitions remains different. 
As a consequence, the LOAF results in the weak-coupling limit discussed above agree with the Bogoliubov and Popov approximations discussed for instance in the Andersen's review article\cite{r:Andersen:2004uq}.
}

%
%
\section{\label{scaled}Rescaled equations}

The above zero-temperature results suggest that $\lambda \rho$ is one of the two characteristic energy scales of the BEC system. At finite temperature, we find the second energy scale is given by $T_0 = 4\pi \gamma [\rho/\zeta(3/2)]^{2/3}$, the critical temperature of the non-interacting Bose gas. It appears that $T_0$ represents the temperature scale of the BEC system, and it is convenient  to supplement the set of scaled variables given in Eq.~\eqref{eq:T0_scaled},  by introducing the scaled temperature,
$
    \tilde T = T / T_0
$.
In this context, we note the useful results $\lambda \rho / T_0 = c \, \xi$ and $(\mathcal{A}/\rho)T_0^{3/2} = 2 \, [\sqrt \pi \, \zeta(3/2)]^{-1}$, with $c=2 \, \zeta^\frac{2}{3}(3/2)$.

In terms of the rescaled variables, we have
\begin{align}
   1
   & = \frac{\rho_0}{\rho}
   + \frac{8}{3\sqrt \pi} \, \xi^{3/2} \, \tilde \Delta^{3/2}
   + \tilde T^{3/2} \,  I_1 (\delta)
   \>,
   \notag \\
\label{eq:gap_scale}
   \frac{\rho_0}{\rho}
   & =
   \tilde \Delta
   + \frac{8}{\sqrt \pi} \, \xi^{3/2} \, \tilde \Delta^{3/2}
   - \tilde T^{3/2} \, I_2 (\delta)
   \>,
\end{align}
with the notations $\delta = \Delta / T  = c \, \xi \, (\tilde \Delta / \tilde T)$, and
\begin{align}
\label{I1_kita}
    I_1(\delta)
    & =
    \frac{2}{\sqrt \pi \, \zeta(3/2) }
    \int_0^\infty
    \frac{\epsilon + \delta}{\sqrt{\epsilon + 2 \delta }} \,
    \frac{d \epsilon}{e^{\sqrt{\epsilon ( \epsilon + 2 \delta )}} - 1}
    \>,
    \\
\label{I2_kita}
    I_2(\delta)
    & =
    \frac{2}{\sqrt \pi \, \zeta(3/2) }
    \int_0^\infty
    \frac{\delta}{\sqrt{\epsilon + 2 \delta }} \,
    \frac{d \epsilon}{e^{\sqrt{\epsilon ( \epsilon + 2 \delta )}} - 1}
    \>,
    \\
\label{I3_kita}
    I_3(\delta)
    & =
    \frac{2}{\sqrt \pi \, \zeta^{\frac{5}{3}}(3/2) }
    \int_0^\infty \!\!
    \sqrt \epsilon \, d \epsilon \,
    \ln ( 1 - e^{- \sqrt{\epsilon ( \epsilon + 2 \delta )}} )
    \>.
\end{align}
The gap equations combine to give
\begin{align}
\label{Delta_bs}
   \tilde \Delta
   & =
   1 - \frac{32}{3 \sqrt \pi} \, \xi^{3/2} \, \tilde \Delta^{3/2}
   - \tilde T^{3/2} \Bigl [ I_1 (\delta) - I_2 (\delta) \Bigr ]
   \>,
   \\
\label{rho0_bs}
   \frac{\rho_0}{\rho}
   & =
   1 - \frac{8}{3\sqrt \pi} \, \xi^{3/2} \, \tilde \Delta^{3/2}
   - \tilde T^{3/2} \,  I_1 (\delta)
   \>,
   \\
\label{p_bs}
   \tilde p / \rho
   & =
   1 - \frac{1}{2} \tilde \Delta^2
   - \frac{64}{15 \sqrt \pi} \, \xi^{3/2} \tilde \Delta^{5/2}
   - \frac{1}{2\xi} \tilde T^{5/2} \, I_3 (\delta)
   \>.
\end{align}
We introduce the rescaled entropy density as
\begin{equation}
    \tilde s = \frac{T_0}{\lambda \rho} \, s
    \>,
\end{equation}
given as, see Eq.~\eqref{eq:gap_s},
\begin{equation}\label{eq:gap_sr}
   \frac{\tilde s}{\rho}
   = \frac{1}{2 \xi} \, \tilde T^{3/2} \, 
   \bigl [  I_4(\delta) - I_3 (\delta) \bigr ]
   \>,
\end{equation}
where we introduced the notation
\begin{equation}
   I_4(\delta)
    =
    \frac{2}{\sqrt \pi \, \zeta^{\frac{5}{3}}(3/2) }
    \int_0^\infty \!\!
    \sqrt \epsilon \, d \epsilon \,
    \frac{\omega(\delta)}{e^{\omega(\delta)} - 1}
    \>.
\end{equation}
The energy density is 
\begin{equation}\label{eq:gap_er}
    \tilde \varepsilon = - \tilde p + \tilde T \tilde s + \tilde \mu \rho
    \>.
\end{equation}

%
%
\section{\label{critical}Critical properties in the weakly-interacting limit}

In the following we will follow closely the approach outlined by Kita in Ref.~\onlinecite{r:Kita:2006kx}. It is important to note that despite the fact that Eqs.~\eqref{Delta_bs}, \eqref{rho0_bs} and \eqref{p_bs} are different from Kita's Eqs.~(41)-(43), in the weakly-interacting limit they become the same. Therefore, in the broken-symmetry phase, for $T < T_c$, our results match closely Kita's results. Differences are simply due to the fact that we found better second-order approximations of the integrals~\eqref{I1_kita}, \eqref{I2_kita} and \eqref{I3_kita}.

In the weakly-interacting limit, we have $\delta \ll 0$ and the temperature-dependent integrals can be approximated as (see App.~\ref{app})
\begin{align}
   I_1 (\delta)
   & \approx
   1 - b_1 \, \delta^{1/2}
      \!\! + b_2 \ \delta
   \>,
   \\
   I_2 (\delta)
   & \approx
      b_1 \, \delta^{1/2}
      - b_2' \ \delta
   \>,
   \\
   I_3(\delta)
   & \approx
    \frac{1}{\zeta^{\frac{2}{3}}(3/2)} \,
    \Bigl ( b_0'
      + \delta
      - \frac{4b_1}{3} \delta^{3/2}
      + \frac{3b_1}{4} \delta^2
   \Bigr )
   \>,
\end{align}
with $b_0' = - \zeta(5/2) / \zeta(3/2)$, $b_1 = \sqrt{2 \pi} / \zeta(3/2)$, $b_2 = b_1 / 2$,  $b_2' = b_1$.
As indicated above, our approximations of the integrals~\eqref{I1_kita}, \eqref{I2_kita} and \eqref{I3_kita} differ from Kita's approximations -- see Eqs. (48a), (48b), and (48c) in Ref.~\onlinecite{r:Kita:2006kx} -- at the second order in the $b_2$ and $b_2'$ coefficients.

Then, the gap equations~\eqref{eq:gap_scale}  read
\begin{align}
   \frac{\rho_0}{\rho}
   &=
   1
   -
   \tilde T^\frac{3}{2} \bigl ( 1 - b_1 x + b_2 x^2 \bigr )
   \>,
   \label{gap_x-chi}
   \\
   \frac{\rho_0}{\rho}
   &=
   \frac{x^2 \tilde T}{c\xi}
   -
   \tilde T^\frac{3}{2} \bigl ( b_1 x - b_2' x^2 \bigr )
   \>,
   \label{gap_x-A}
\end{align}
where we introduced the variable $x^2 = \delta = c \, \xi \, (\tilde \Delta / \tilde T)$, with $c=2 \, \zeta^\frac{2}{3}(3/2)$.
Combining Eqs.~\eqref{gap_x-chi} and~\eqref{gap_x-A},  we obtain
\begin{align}
   x(\tilde T) = & \frac{b_1 (c \xi) \tilde T^\frac{1}{2}}{1 + (b_2+b_2') (c \xi) \tilde T^\frac{1}{2}}
   \label{x_of_T}
   \\ \notag & \times
   \Biggl \{ 1 + \sqrt{ 1 - \frac{\tilde T^\frac{3}{2} - 1}{b_1^2 (c \xi) \tilde T^2}
                      \Bigl [ 1 + (b_2+b_2') (c \xi) \tilde T^\frac{1}{2} \Bigr ] }
   \Biggr \}
   \>,
\end{align}
and we can write Eq.~\eqref{gap_x-A} as
\begin{align}
   \frac{\rho_0}{\rho}
   = \frac{1 + b_2' (c \xi) \tilde T^\frac{1}{2}}{c \xi} \tilde T x
   \Bigl [ x - \frac{b_1 (c \xi) \tilde T^\frac{1}{2}}{1 + b_2' (c \xi) \tilde T^\frac{1}{2}}
   \Bigr ]
   \>.
   \label{rho0_of_T}
\end{align}
At the critical point we have $\rho_0=0$.
Then, from Eq.~\eqref{rho0_of_T}, we obtain
\begin{equation}
\label{xc}
   x_c \equiv
   x(\tilde T_c) = \frac{b_1 (c \xi) \tilde T_c^\frac{1}{2}}{1 + b_2' (c \xi) \tilde T_c^\frac{1}{2}}
   \>.
\end{equation}
The expansion of  $\tilde T_c$ in powers of $\xi$ is obtained from Eqs.~\eqref{x_of_T} and \eqref{xc} as
\begin{align}
    \tilde T_c = & 1 + \frac{2}{3} b_1^2 c \, \xi
    + \frac{2}{3} b_1^2 c \Bigr [ \frac{7}{6} b_1^2 c - (b_2 + b_2') c \Bigr ] \, \xi^2 + \cdots
    \label{T_c}
    \>.
\end{align}
which gives
\begin{align}
    \frac{T_c - T_0}{T_0}
    = &
    \frac{8\pi}{\zeta^\frac{4}{3}(3/2)} \, \xi
    \label{Tc:DeltaTc}
    \\ \notag &
    +
    \frac{8\pi}{\zeta^\frac{4}{3}(3/2)}
    \biggl [ \frac{14\pi}{3\zeta^\frac{4}{3}(3/2)} - \frac{3\sqrt{2\pi}}{\zeta^\frac{1}{3}(3/2)} \biggr ] \, \xi^2
    + \cdots \>.
\end{align}
Here, the linear coefficient is $\approx 2.33$, whereas the quadratic coefficient is
$\approx - 3.23$.
The result for the linear coefficient in $\xi$ is the same as that obtain by Baym \emph{et al.} using the large-$N$ expansion for the critical theory\cite{PhysRevLett.83.1703,r:Baym:2000fk}. Kita also obtained this linear coefficient, see Eq.~(52) in Ref.~\onlinecite{r:Kita:2006kx}.

\textcolor{black}{We note that the large-$N$ expansion results obtained by Baym \emph{et al.}  for the critical theory\cite{PhysRevLett.83.1703,r:Baym:2000fk} were later improved by Kleinert\cite{kleinert:2003mplb} and Kastening\cite{PhysRevA.68.061601,PhysRevA.69.043613} using five-, six- and seven-loop variational perturbation theory, respectively. At the seven-loop order, Kastening\cite{PhysRevA.69.043613} calculated a value $(T_c - T_0)/T_0 = (1.27 \pm 0.11) \xi$, which is in excellent agreement with Monte Carlo lattice field-theory results\cite{PhysRevLett.87.120401,PhysRevLett.87.120402,PhysRevE.64.066113}.  The quadratic coefficient was calculated by Arnold, Moore and Tomasik\cite{PhysRevA.65.013606}, yielding also a second-order logarithmic correction that cannot be captured by LOAF, which is only a one-loop approximation. Their result, $(T_c - T_0)/T_0 \approx (1.32 \pm 0.02)\xi + [ 19.7518 \, \ln(\xi) + (75.7 \pm 0.4)] \xi^2$, indicates that two-loop contributions are also important in determining the value of the quadratic coefficient, as the LOAF result  is too small.
}

%
\begin{figure}[t]
   \centering
   \includegraphics[width=0.9\columnwidth]{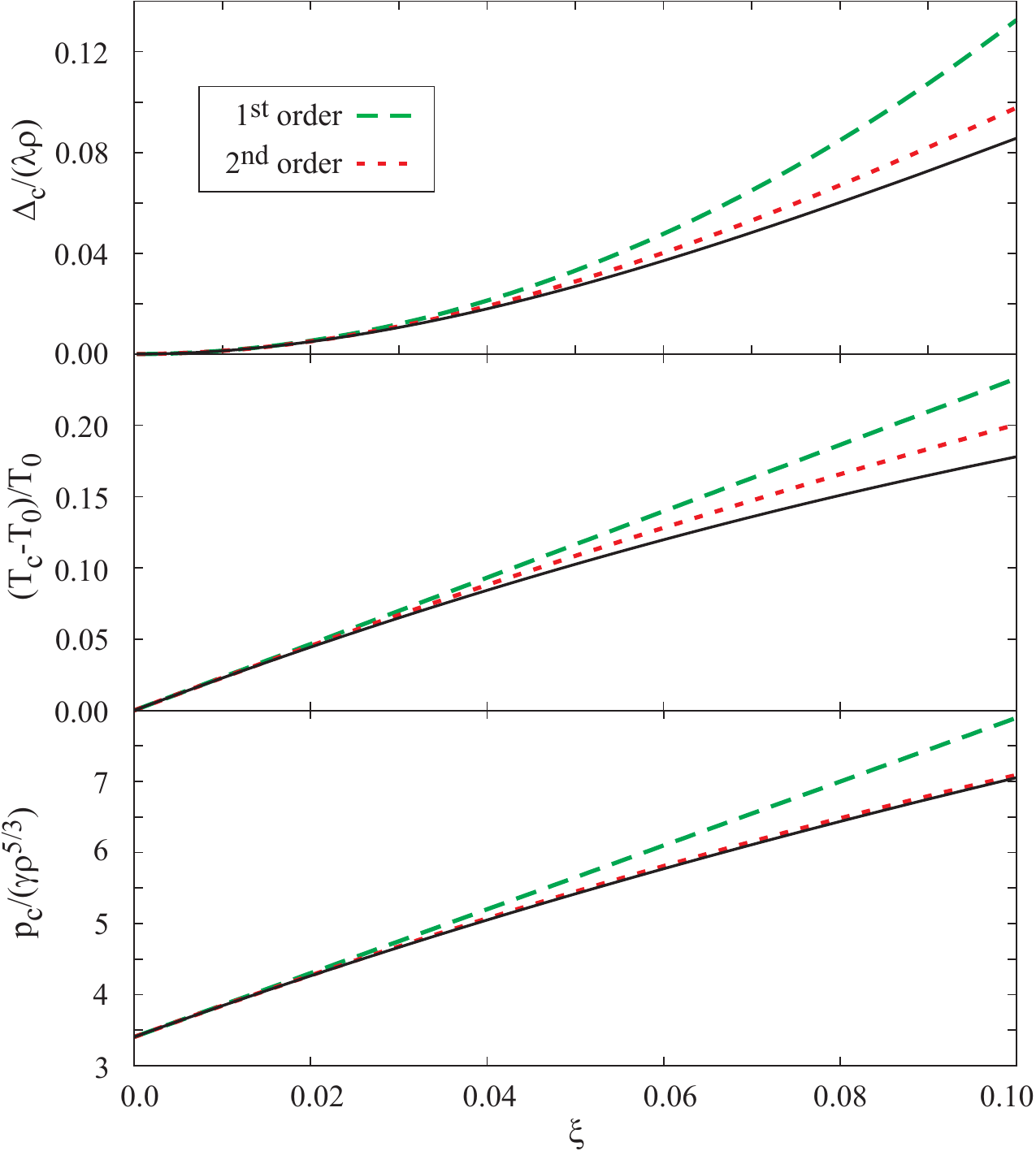}
   \caption{\label{Fig_Tc_weak}(Color online) Comparisons of the exact critical values of the auxiliary field, $\Delta_c$, the ratio $(T_c - T_0)/T_0$, and pressure, $p_c$, and their respective first- and second-order approximations,  as a function of $\xi = \rho^{1/3}a_0$, in the weakly-interacting regime.
   The LOAF approximations for the critical values of the auxiliary field, $\Delta_c$, the ratio $(T_c - T_0)/T_0$, and pressure, $p_c$, in the weakly-interacting regime, are given in Eqs.~\eqref{Tc:Deltac}, \eqref{Tc:DeltaTc} and \eqref{Tc:pc}, respectively.
   }
\end{figure}
%

Substituting $\Tilde T_c$ from Eq.~\eqref{T_c} in Eq.~\eqref{xc}, we obtain
\begin{equation}
    x_c
    = b_1 c \, \xi
    + b_1 c \, \Bigl ( \frac{1}{3} b_1^2 c - b_2' c \Bigr ) \, \xi^2
    + \cdots
    \>.
\end{equation}
and
\begin{equation}
    \frac{\Delta_c}{T_0}
    = b_1^2 c^2 \, \xi^2
    + b_1^2 c^2 \, \Bigl ( \frac{4}{3} b_1^2 c - 2 b_2' c \Bigr ) \, \xi^3
    + \cdots
    \>.
    \label{Tc:Deltac}
\end{equation}
In the latter, the quadratic coefficient is $8\pi / \zeta^\frac{2}{3}(3/2) \approx 13.25$, whereas the coefficient of the cubic term is $\approx -34.76$.
The leading-order approximation was also obtained by Kita, see Eq.~(53) in Ref.~\onlinecite{r:Kita:2006kx}.

The temperature dependence of $\Delta$ close to $T_c$ is derived from
\begin{align}
\label{xxc}
   & x(\tilde T) - x(\tilde T_c)
   \\ \notag & =
   - b_1 b_2 c^2 \, \xi + \sqrt{b_1^2 c^2 \, \xi^2 + \frac{3}{2} c \, \xi \Bigl ( 1 - \frac{\tilde T}{\tilde T_c} \Bigl )}
   + \cdots
   \>,
\end{align}
as
\begin{align}
\label{TTc}
   & \frac{\Delta(T)}{T_0}  - \frac{\Delta(T_c)}{T_0}
   \\ \notag & =
   b_1^2 c^2 \, \xi^2 + \sqrt{4 b_1^4 c^4 \, \xi^4 + 6 b_1^2 c^3 \, \xi^3 \Bigl ( 1 - \frac{T}{T_c} \Bigl )}
   + \cdots
   \>.
\end{align}
The Eq.~\eqref{xxc} above is the same as Kita's Eq.~54.
Similarly, we obtain
\begin{align}
\label{rho0c}
   \frac{\rho_0}{\rho}
   =
   b_1^2 c \, \xi + \sqrt{b_1^4 c^2 \, \xi^2 + \frac{3}{2} b_1^2 c \, \xi \Bigl ( 1 - \frac{T}{T_c} \Bigl )}
   + \cdots
   \>.
\end{align}

The critical pressure in the weakly-interacting limit is obtained from
\begin{equation}
   \tilde p_c / \rho
   \approx
   1
   -
   \frac{\tilde T_c^{5/2}}{2 \xi \zeta^{\frac{2}{3}}(3/2)} \,
   \Bigl (
    b_0'
      + x
      - \frac{4b_1}{3} x^3
      + \frac{3b_1}{4} x^4
   \Bigr )
   \>.
\end{equation}
This gives
\begin{align}
   & \tilde p_c / \rho
   =
   \frac{\zeta(5/2)}{2 \zeta^{\frac{5}{3}}(3/2)} \frac{1}{\xi}
   +
   \Bigl [
   1 +
   \frac{5}{6} \, \frac{\zeta(5/2)}{\zeta^{\frac{5}{3}}(3/2)}
   b_1^2 c
   \Bigr ] 
   \label{Tc:pc}
   \\ \notag &
   - \frac{b_1^2 c^2}{2\zeta^{\frac{2}{3}}(3/2)} \,
   \Bigl \{
   1 - \frac{5}{3} \, \frac{\zeta(5/2)}{\zeta(3/2)} \, \Bigl [ \frac{5}{3} b_1^2 - (b_2 + b_2') \Bigr ]
   \Bigr \} \, \xi
   + \cdots \>.
\end{align}
In the noninteracting limit, we obtain $p_c(\Tilde T_0)/(\gamma \rho^{5/3}) = 4\pi \, \zeta(5/2) / \zeta^{\frac{5}{3}}(3/2) \approx 3.4$. The coefficient of the linear term is $8\pi \bigl [ 1 + (10 \pi / 3) \,  \zeta(5/2) / \zeta^{3}(3/2) \bigr ] \approx 44.94$, whereas the coefficient of the quadratic contribution is $\approx - 6.42$.

For illustrative purposes, in Fig.~\ref{Fig_Tc_weak} we depict the exact critical values of the auxiliary field, $\Delta_c$, the ratio $(T_c - T_0)/T_0$, and pressure, $p_c$, and their respective first- and second-order approximations,  as a function of $\xi = \rho^{1/3}a_0$, in the weakly-interacting regime.

%
\begin{figure}[t]
   \centering
   \includegraphics[width=0.9\columnwidth]{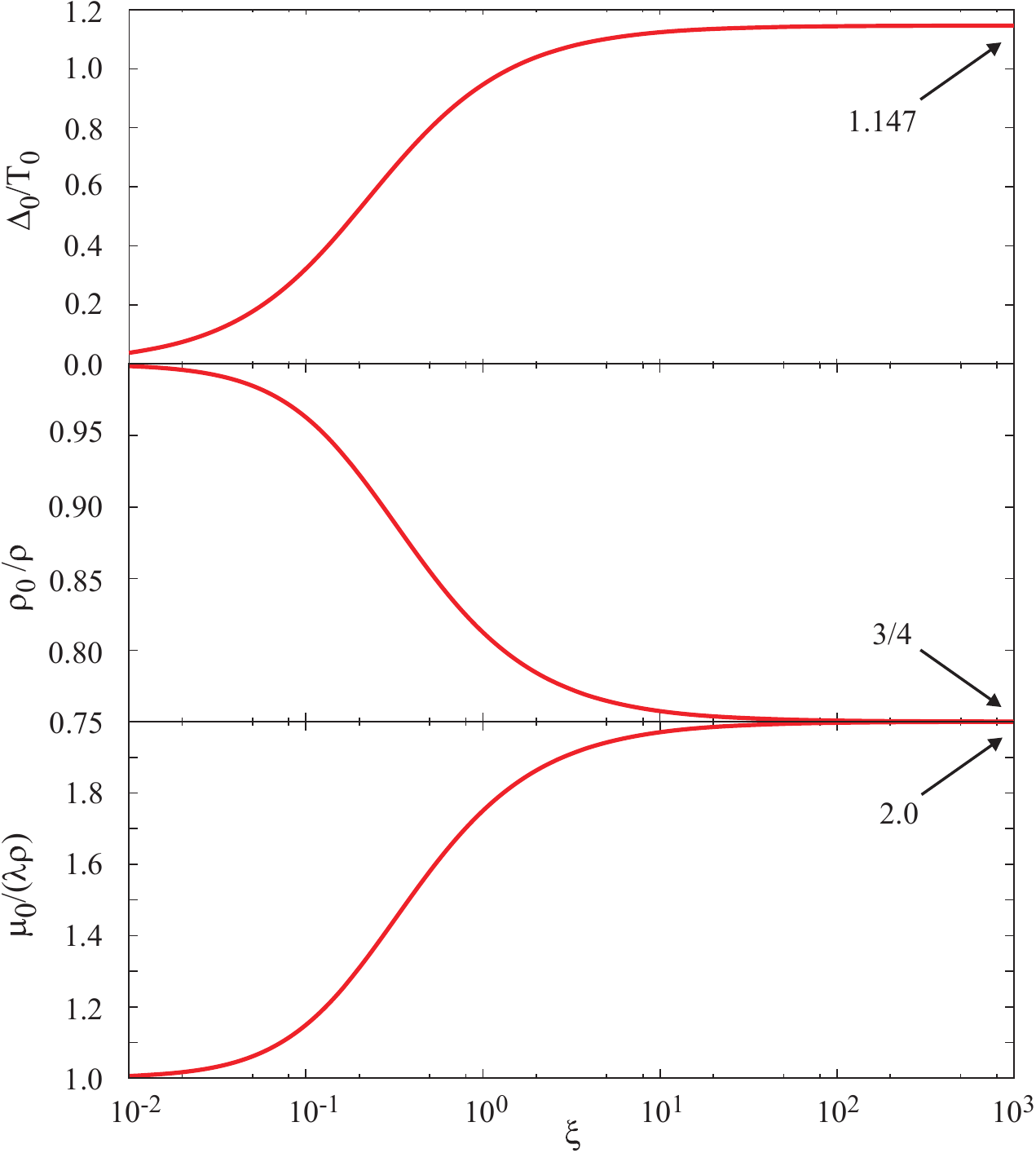}
   \caption{\label{Fig_T0_params} (Color online) Zero-temperature values of the auxiliary field, $\Delta_0$, condensate fraction, $\rho_0/\rho$,  and chemical potential, $\mu_0$, as a function of  $\xi = \rho^{1/3}a_0$. The condensate fraction, $\rho_0/\rho$,  and chemical potential, $\mu_0/(\lambda \rho)$, plots depicted in the bottom two panels, were first shown in Fig.~2 of Ref.~\onlinecite{PhysRevA.Bose:2011}. The reader is directed to Ref.~\onlinecite{PhysRevA.Bose:2011} for further studies and derivations. Here, we emphasize that the numerical values of the  auxiliary field, $\Delta_0$, condensate fraction, $\rho_0/\rho$,  and chemical potential $\mu_0$, in the unitarity limit, compare well with the exact solutions given in Eq.~\eqref{unit:Delta0}, \eqref{unit:rho0} and Eq.~\eqref{unit:mu}, respectively.
   }
\end{figure}
%
\begin{figure}[t]
   \centering
   \includegraphics[width=0.9\columnwidth]{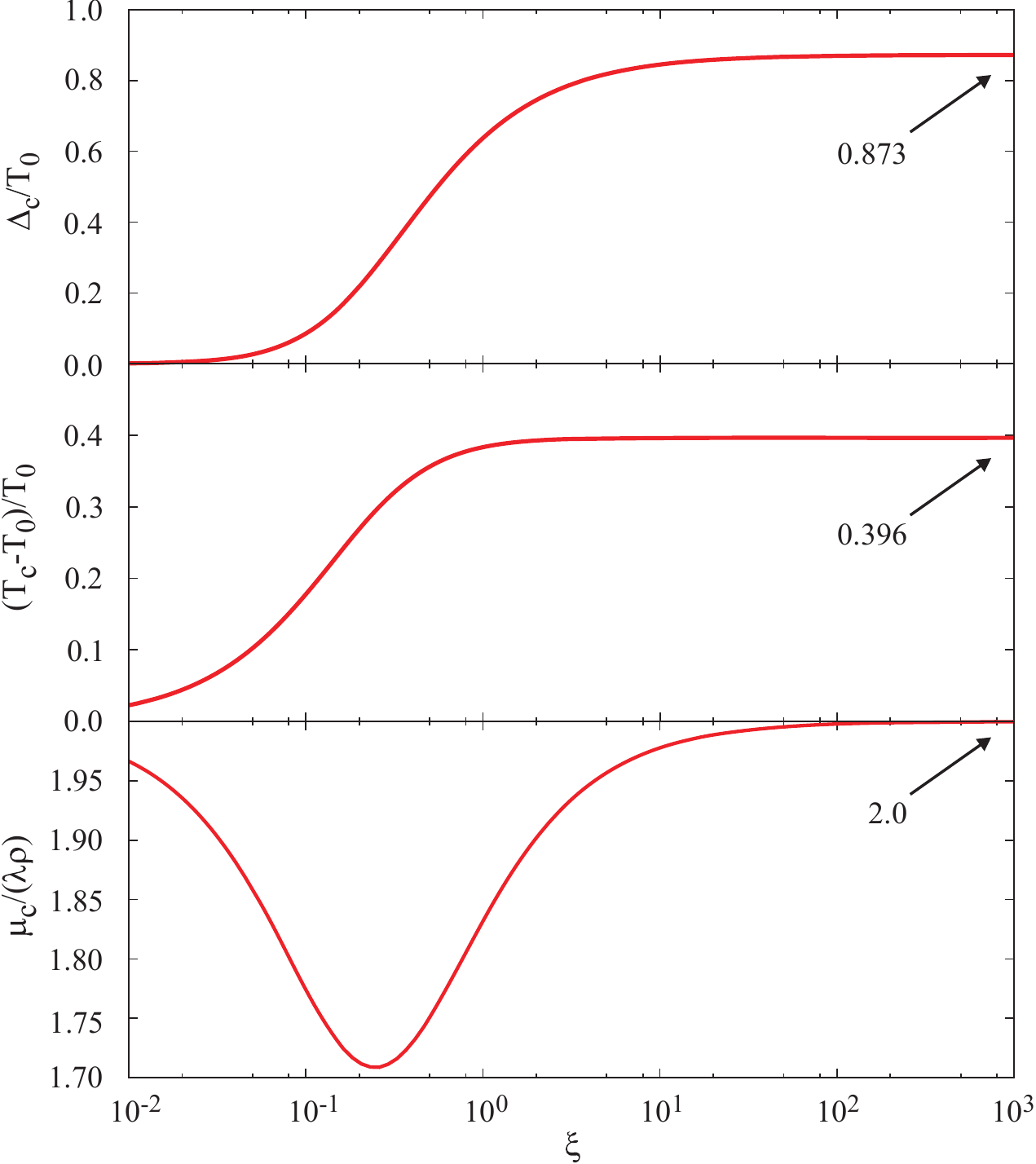}
   \caption{\label{Fig_Tc_params} (Color online) Critical values of the auxiliary field, $\Delta_c$,  the ratio $(T_c - T_0)/T_0$, and chemical potential, $\mu_c$, as a function of  $\xi = \rho^{1/3}a_0$.
The critical auxiliary field, $\Delta_c$,  and the ratio $(T_c - T_0)/T_0$, depicted in the top two panels, were reported previously in Fig.~6 of Ref.~\onlinecite{PhysRevA.Bose:2011}, and further discussions regarding these plots can be found in Ref.~\onlinecite{PhysRevA.Bose:2011}. Here, we emphasize that the numerical values of the critical auxiliary field, $\Delta_c$, and the ratio $(T_c - T_0)/T_0$,  in the unitarity limit, compare well with the semi-analytical approximations given in Table~1. The numerical value of the chemical potential $\mu_c$, agrees with the exact result obtained in  Eq.~\eqref{unit:mu}.
   }
\end{figure}
%

%
%
\section{\label{unitarity}Critical properties in the unitarity limit}

The LOAF approximation is a non-perturbative approximation. Therefore one can solve Eqs.~\eqref{Delta_bs} and \eqref{rho0_bs} for arbitrary values of coupling constant related to the dimensionless parameter, $\xi = \rho^{1/3} a_0$, to calculate the values of the condensate fraction, auxiliary field, $\Delta$, and all thermodynamical variables derivable from the grand potential related to the pressure~\eqref{p_bs}.   For illustrative purposes, in Fig.~\ref{Fig_T0_params} we depict the zero-temperature values of the auxiliary field $\Delta_0$, condensate fraction, $\rho_0/\rho$,  and chemical potential $\mu_0$, as a function of  $\xi$, whereas in Fig.~\ref{Fig_Tc_params} we show the dependence of the critical values of the auxiliary field, $\Delta_c$,  the ratio $(T_c - T_0)/T_0$, and chemical potential, $\mu_c$, as a function of  $\xi = \rho^{1/3}a_0$. From Fig.~\ref{Fig_Tc_params} we conclude that in the unitarity limit (i.e. in the limit $\xi$~$\rightarrow$~$\infty$), we have $\Delta_c /T_0 \rightarrow 0.873$ and $T_c /T_0 \rightarrow 1.396$.

In the previous sections we focused on analytic approximations of the solutions to the gap equations, Eqs.~\eqref{Delta_bs} and \eqref{rho0_bs}, for the weakly-interacting limit. These approximations were derived by dropping the terms proportional to $\tilde \Delta^{3/2}$ and higher in Eqs.~\eqref{Delta_bs}, \eqref{rho0_bs} and~\eqref{p_bs}, and by approximating the integrals~\eqref{I1_kita}, \eqref{I2_kita} and \eqref{I3_kita} as described in App.~\ref{app}. This approach is made possible by the fact that the critical value of the auxiliary field, $\tilde \Delta$ is small in the weakly-interacting limit, and $\tilde \Delta \ll 1$ leads to $\delta = \tilde \Delta / \tilde T \ll 1$ as $\tilde T > 1$. 

A similar approach is possible for obtaining analytical approximations for the critical values of the auxiliary field, $\Delta_c$, and the critical temperature $T_c$, in the unitarity limit. This approach requires improving the approximations to the integrals~\eqref{I1_kita}, \eqref{I2_kita} and \eqref{I3_kita} described in App.~\ref{app}, by supplementing the approximation of $I_2(\delta)$ by the term $(19/64)b_1 \delta^{3/2}$, and by adding the term $- (99/256) b_1 \delta^{3/2}$ to the approximation of $I_1(\delta)-I_2(\delta)$. This allows us to take into account the  terms proportional to $\tilde \Delta^{3/2}$ in Eqs.~\eqref{Delta_bs}, \eqref{rho0_bs} and~\eqref{p_bs}. We will describe this procedure next.

As already discussed, the unitarity limit corresponds to the strongly-interacting limit, $\xi \rightarrow \infty$. In this regime, the energy scale $\lambda \rho$ diverges, and the only scale remaining in the problem is the temperature scale, $T_0$. Hence, in the unitarity limit the gap equations are scaled by introducing the variable, $\bar \Delta = \Delta / T_0$. From Eqs.~\eqref{eq:gap_bs1} and \eqref{eq:gap_bs2}, we obtain
\begin{align}
\label{eq:Delta_unit}
   1
   & =
   \frac{\rho_0}{\rho} 
   +
   \frac{2}{3} \sqrt{ \frac{2}{\pi} } \, \frac{\bar \Delta^{3/2}}{\zeta(3/2)}
   + \tilde T^{3/2} \,  I_1 (\delta)
   \>,
   \\
\label{eq:rho0_unit}
   \frac{\rho_0}{\rho}
   & =
   2 \, \sqrt{ \frac{2}{\pi} } \, \frac{\bar \Delta^{3/2}}{\zeta(3/2)}
   - \tilde T^{3/2} \, I_2 (\delta)
   \>,
\end{align}
with $\delta = \Delta / T = \bar \Delta / \tilde T$. 
By adding the gap equations, we obtain also
\begin{align}
\label{eq:T_unit}
   1
   =
   \frac{8}{3} \sqrt{ \frac{2}{\pi} } \, \frac{\bar \Delta^{3/2}}{\zeta(3/2)}
   + \tilde T^{3/2} \bigl [ I_1(\delta) - I_2(\delta) \bigr ]
   \>.
\end{align}

At zero temperature, $T=0$, we can solve \eqref{eq:T_unit} to obtain the value of the auxiliary field in the unitarity limit (UL) at zero temperature, as
\begin{equation}
    \Delta_0 / T_0 = \Bigl [ \, \frac{3}{8} \sqrt{ \frac{\pi}{2} } \, \zeta(3/2) \, \Bigr ]^{2/3}
    \approx 1.147
    \label{unit:Delta0}
    \>.
\end{equation}
and from ~\eqref{eq:rho0_unit} we obtain the UL condensate fraction is
\begin{equation}
    \frac{\rho_0}{\rho} = \frac{3}{4} \>.
    \label{unit:rho0}
\end{equation}
Alternatively, the UL condensate depletion fraction is~1/4.

The critical temperature, $\delta_c$ in the unitarity limit is obtained by solving  Eq.~\eqref{eq:rho0_unit} with $\rho_0=0$. Assuming the expansion of $I_2(\delta)$ as a series powers of $\delta$ is truncated by dropping terms proportional to $\delta^{3/2}$ and higher, we find $\delta_c$ as the solution of a quadratic equation. Then we can calculate $\tilde T_c = T_c/T_0$ from Eq.~\eqref{eq:T_unit}. Table~\ref{table_unit} summarizes approximations for the critical values of the auxiliary field, $\Delta_c$, and the critical temperature $T_c$, in the unitarity limit. These approximations are compared with the numerical ``exact'' values. The errors of the third-order approximations relative to the exact values are less than half of a percent. 

Finally, we can show that the equation of state in the unitarity limit is independent of temperature. We begin, by calculating the UL asymptotes of the chemical potential and pressure. For arbitrary  temperature, Eqs.~\eqref{eq:mu_r0} and~\eqref{p_bs}  give
\begin{equation}
    \mu = 2 \, \lambda \rho
    \>,
    \label{unit:mu}
\qquad
   p = \lambda \rho^2
    \>.
\end{equation}
Then, from Eq.~\eqref{eq:E_r0} we find the UL equation of state at zero temperature is
\begin{equation}
    E_0 / (p_0 V) = \varepsilon_0 / p_0 = \tilde \varepsilon_0 / \tilde p_0 = 1
    \>.
\end{equation}
For finite temperature we use instead Eqs.~\eqref{eq:gap_er} and~\eqref{eq:gap_sr} and obtain
\begin{equation}
    E / (p V) = \varepsilon / p = \tilde \varepsilon / \tilde p = 1
    \>.
\end{equation}
This result is different than in the Fermi gas case, where the equation of state in the unitarity limit is,  $E / (p V)$$=$$3/2$ ~\cite {PhysRevA.Fermi:2011,PhysRevLett.92.090402}.

%
%
\begin{table}[t]
\caption{\label{table_unit}
Exact and approximate values of the critical values of the auxiliary field, $\Delta_c$, and the critical temperature, $T_c$, in the unitarity limit. 
\\}
\begin{tabular}{lrccc}
\hline
& Exact & 1$^\mathrm{st}$ order & 2$^\mathrm{nd}$ order & 3$^\mathrm{rd}$ order
\\
\hline
$\sqrt{\delta_c}$  & 0.790 & 1.253 & 0.694 & 0.789
\\
$\Delta_c/T_0$    & 0.873 & 4.619 & 0.653 & 0.875
\\
$T_c/T_0$            & 1.396 & 2.941 & 1.356 & 1.407
\\
\hline
\end{tabular}
\end{table}
%
%

%
%
\section{\label{concl}Conclusions}

In summary, in this paper we discussed analytical approximations of the properties of dilute Bose gases using the LOAF approximation in the weakly-interacting and the strongly-interacting (unitarity) limit.  We focus deliberately on the case of Bose gases in the broken-symmetry phase, in order to make contact with analytical results already existing in the literature. The weakly-interacting results at zero-temperature are shown to be identical with those found in the weakly interacting Bose gas theory discussed by Fetter and Walecka\cite{r:Fetter:1971fk}, whereas close to the critical temperature, the LOAF results are similar to those obtained by Kita using the related Luttinger-Ward functional\cite{r:Kita:2005fk,r:Kita:2005uq,r:Kita:2006kx}. In obtaining our results, we have improved the analytical approximation of temperature-dependent integrals, first discussed by Kita\cite{r:Kita:2006kx}. These approximations were then applied to the analytical study of the LOAF predictions in the unitarity limit and found to give good agreement with our numerical results. LOAF predicts that the equation of state for the Bose gas in the unitarity limit is $E / (p V) = 1$, unlike the case of the Fermi gas when $E / (p V) = 3/2$.


\acknowledgements

Work performed in part under the auspices of the U.S. Department of Energy.
The authors would like to thank E. Mottola for useful discussions and the Santa Fe Institute for its hospitality during this work.

%
\appendix
%
%
\section{\label{app}Approximations of certain integrals}

For completeness, in this appendix we will derive the first- and second-order approximations of the integrals~\eqref{I1_kita}, \eqref{I2_kita} and \eqref{I3_kita}. Our approach follows closely the discussion in Kita's paper, see Ref.~\onlinecite{r:Kita:2006kx}. Our first-order approximations are the same as Kita's, but we differ at the second order.

Consider the integral
\begin{equation}
      I_1(z)
      =
    \frac{2}{\sqrt \pi \, \zeta(3/2) }
      \int_0^\infty
      \frac{\epsilon + z}{\omega} \,
      \frac{\sqrt{\epsilon} \ d \epsilon}{e^\omega - 1}
      \>,
\end{equation}
with $\omega = \sqrt{\epsilon ( \epsilon + 2 z )}$.
In the weakly-coupling limit we have $z \rightarrow 0$, as $\lambda \rightarrow 0$, and we seek a power expansion of $I_1(z)$ in powers of $z$. We have
\begin{equation}
      \frac{\epsilon + z}{\omega} \,
      \frac{\sqrt{\epsilon}}{e^\omega - 1}
      =
      \frac{\sqrt{\epsilon}}{e^\epsilon - 1}
      -
      \frac{\sqrt{\epsilon} \ e^\epsilon}{(e^\epsilon - 1)^2}
      + \mathcal{O}(\epsilon^2)
      \>.
\end{equation}
Upon integration with respect to $\epsilon$, we find that only the first term converges,
\begin{equation}
    b_0 =
    \frac{2}{\sqrt \pi \, \zeta(3/2) }
      \int_0^\infty
      \frac{\sqrt{\epsilon} \ d \epsilon}{e^\epsilon - 1}
     = 1
     \>.
\end{equation}
Therefore, we can write $I_1(z) = b_0 + I_{1,\infty}(z)$, with the remainder
\begin{equation}
      I_{1,\infty}(z)
      =
    \frac{2}{\sqrt \pi \, \zeta(3/2) }
      \int_0^\infty \!\! d \epsilon \,
      \Bigl (
      \frac{\epsilon + z}{\omega} \,
      \frac{\sqrt{\epsilon}}{e^\omega - 1}
      -
      \frac{\sqrt{\epsilon}}{e^\epsilon - 1}
      \Bigr )
\label{eq:I1_inf}
      \>.
\end{equation}
The leading-order approximation of $ I_{1,\infty}(z)$ is obtained by expanding out the exponentials in the denominator to first order. We obtain
\begin{equation}
     I_{1,\infty}^{(1)}(z)
     =
    \frac{2}{\sqrt \pi \, \zeta(3/2) }
      \int_0^\infty \!
      \Bigl (
      \frac{\epsilon + z}{\epsilon + 2 z}
      - 1
      \Bigr )
      \frac{d \epsilon}{\sqrt{\epsilon}}
      =
      - b_1 \sqrt{z}
      \>,
\end{equation}
with $b_1 = \sqrt{2 \pi} / \zeta(3/2)$.
Note that it is important to treat on equal footing the two exponentials in Eq.~\eqref{eq:I1_inf} in order to obtain the correct divergence subtraction.

%
\begin{figure}[b]
   \centering
   \includegraphics[width=0.9\columnwidth]{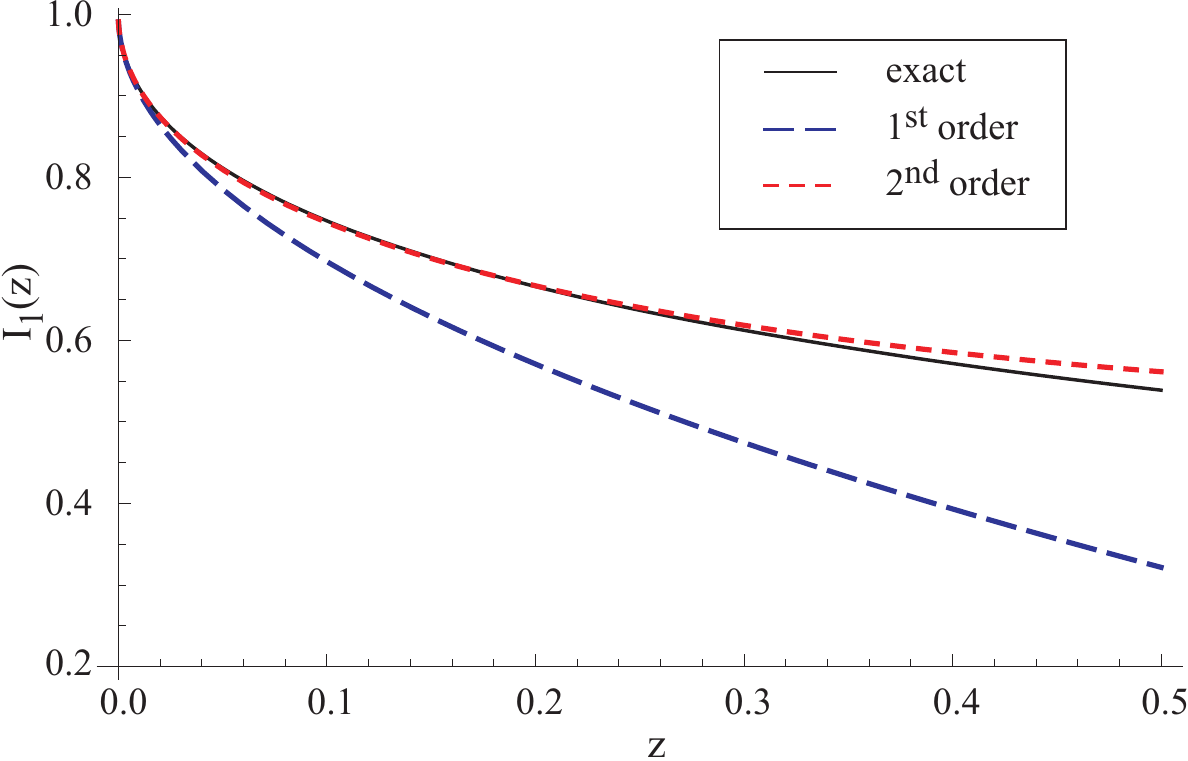}
   \caption{\label{Fig_I1} (Color online) Comparisons of the exact $z$ dependence of $I_1(z)$, and the first- and second-order approximations in powers of~$z$.}
\end{figure}
%
\begin{figure}[t]
   \centering
   \includegraphics[width=0.9\columnwidth]{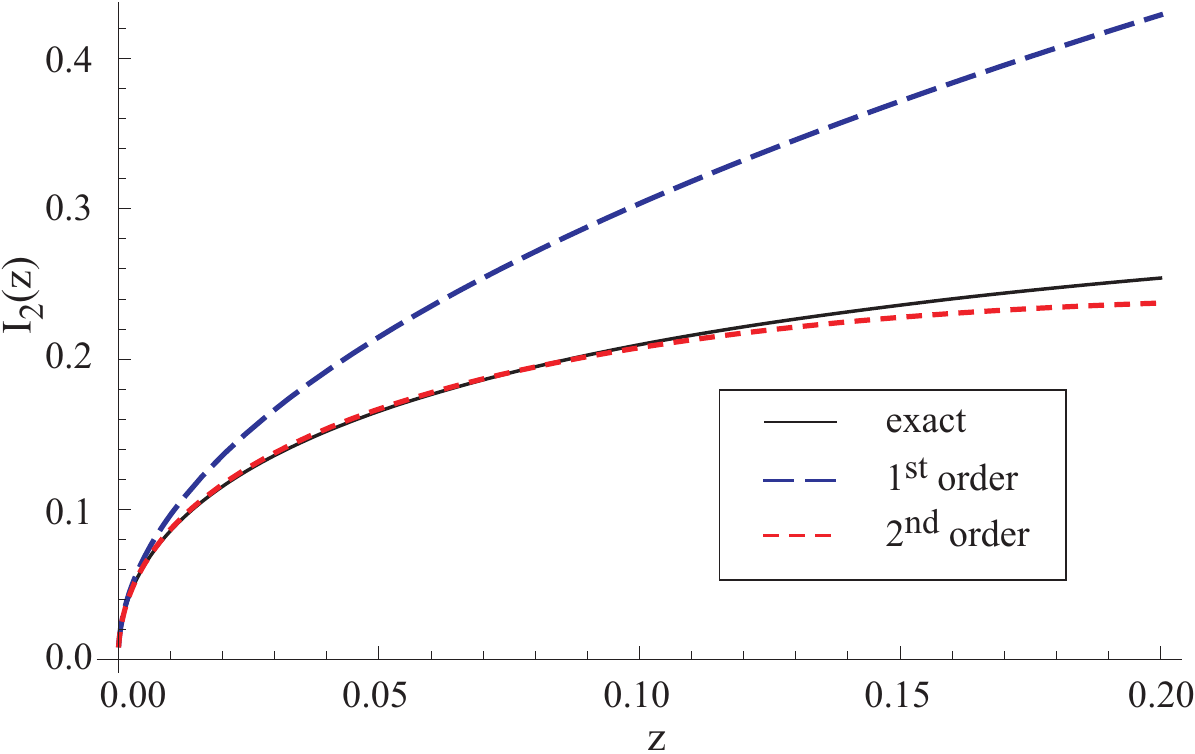}
   \caption{\label{Fig_I2} (Color online) Comparisons of the exact $z$ dependence of $I_2(z)$, and the first- and second-order approximations in powers of~$z$.}
\end{figure}
%
\begin{figure}[t]
   \centering
   \includegraphics[width=0.9\columnwidth]{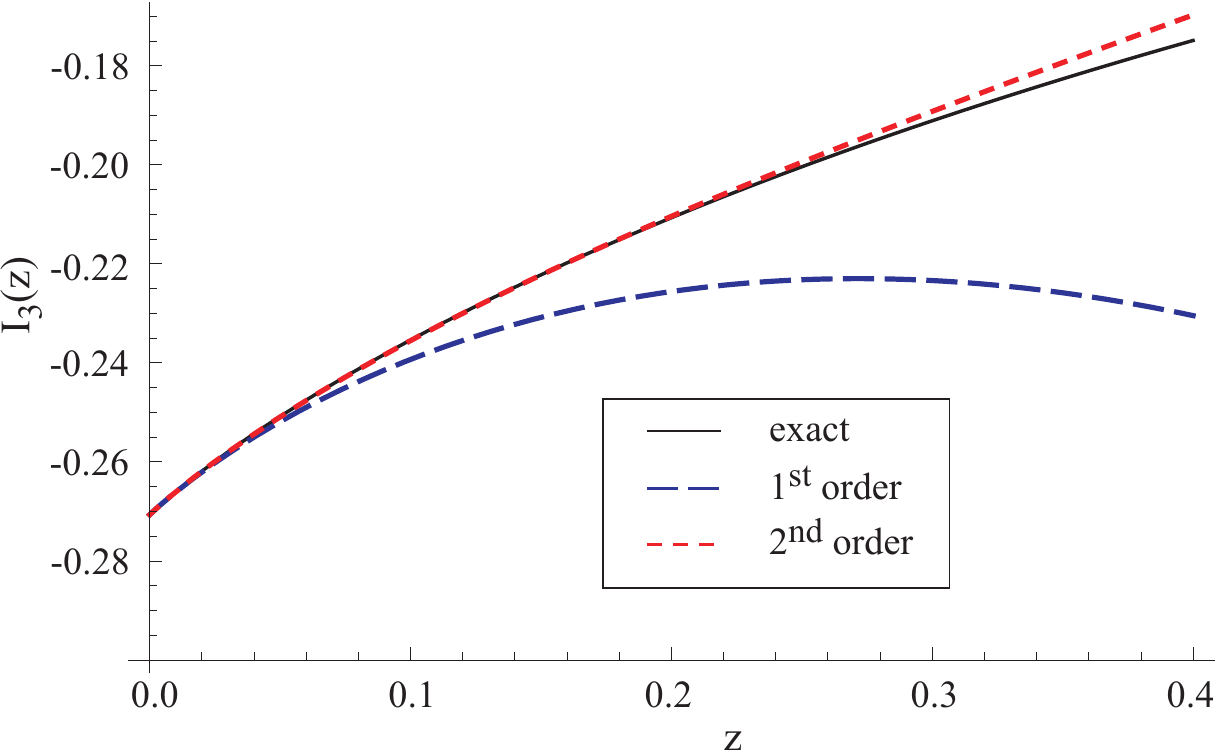}
   \caption{\label{Fig_I3} (Color online) Comparisons of the exact $z$ dependence of $I_3(z)$, and the first- and second-order approximations in powers of~$z$.}
\end{figure}
%

The second-order correction to $I_{1,\infty}(z)$ corresponds to the term linear in $z$ in the power expansion of
\begin{equation}
      \frac{\epsilon + z}{\epsilon + 2 z} \,
      \frac{1}{\sqrt{\epsilon}} \,
      \Bigl ( \,
      \frac{1}{1+\frac{1}{2}\sqrt{\epsilon(\epsilon + 2 z)}}
      - 1 \Bigr )
      \>.
\end{equation}
This gives
\begin{equation}
     I_{1,\infty}^{(2)}(z)
     =
    \frac{2}{\sqrt \pi \, \zeta(3/2) }
      \int_0^\infty
      \frac{z \, \sqrt{\epsilon}}{(\epsilon + 2)^2}
      d \epsilon
      =
      b_2 \ z
      \>,
\end{equation}
with $b_2 = b_1/ 2$.
Therefore, the second-order approximation of $I_1(z)$ is
\begin{equation}
      I_1(z)
      =
    \frac{2}{\sqrt \pi \, \zeta(3/2) } \!
      \int_0^\infty \!\!
      \frac{\epsilon + z}{\omega} \,
      \frac{\sqrt{\epsilon} \ d \epsilon}{e^\omega - 1}
      \approx
      1
      - b_1 \, \sqrt z
      + b_2 \ z
      \>.
\end{equation}
The first- and second-order approximations of $I_1(z)$ are illustrated in Fig.~\ref{Fig_I1}.
The coefficient $b_2=0.559$ calculated by Kita\cite{r:Kita:2006kx} results in a worse second-order approximations.

Similarly, we can evaluate the expansion in powers of $z$ of the integral
\begin{equation}
      I_2(z)
      =
    \frac{2}{\sqrt \pi \, \zeta(3/2) }
      \int_0^\infty
      \frac{z}{\omega} \,
      \frac{\sqrt{\epsilon} \ d \epsilon}{e^\omega - 1}
      \>.
\end{equation}
In this case, there is no analytic part identified after performing the $z$ expansion of the integrand in $I_2(z)$. Therefore, we have $I_2(z) = I_{2,\infty}(z)$, with the remainder
\begin{equation}
      I_{2,\infty}(z)
      =
    \frac{2}{\sqrt \pi \, \zeta(3/2) }
      \int_0^\infty
      \frac{z}{\sqrt{\epsilon + 2 z}} \,
      \frac{d \epsilon}{e^\omega - 1}
\label{eq:I2_inf}
      \>.
\end{equation}
The first-order approximation of $ I_{2,\infty}(z)$ is obtained by expanding out the exponential in the denominator to first order. We obtain
\begin{equation}
     I_{2,\infty}^{(1)}(z)
     =
    \frac{2}{\sqrt \pi \, \zeta(3/2) }
      \int_0^\infty
      \frac{z}{\epsilon + 2 z}
      \frac{d \epsilon}{\sqrt{\epsilon}}
      =
      b_1' \sqrt{z}
      \>,
\end{equation}
with $b_1' = b_1$.
The second-order correction to $I_{2,\infty}(z)$ corresponds to the term linear in $z$ in the power expansion of
\begin{equation}
      \frac{z}{\epsilon + 2 z} \,
      \frac{1}{\sqrt{\epsilon}} \,
      \Bigl ( \,
      \frac{1}{1+\frac{1}{2}\sqrt{\epsilon(\epsilon + 2 z)}}
      - 1 \Bigr )
      \>.
\end{equation}
This gives
\begin{equation}
     I_{2,\infty}^{(2)}(z)
     =
     -
    \frac{2}{\sqrt \pi \, \zeta(3/2) }
      \int_0^\infty
      \frac{z}{\epsilon + 2}
      \frac{d \epsilon}{\sqrt{\epsilon}}
      =
      - \, b_2' \ z
      \>,
\end{equation}
with $b_2' = b_1$.
Therefore, the second-order approximation of $I_2(z)$ is
\begin{equation}
      I_2(z)
      =
    \frac{2}{\sqrt \pi \, \zeta(3/2) }
      \int_0^\infty
      \frac{z}{\omega} \,
      \frac{\sqrt{\epsilon} \ d \epsilon}{e^\omega - 1}
      \approx
      b_1 \, \sqrt z
      - \, b_2' \ z
      \>.
\end{equation}
The first- and second-order approximations of $I_2(z)$ are illustrated in Fig.~\ref{Fig_I2}.
The coefficient $b_2'=1.12$ calculated by Kita\cite{r:Kita:2006kx} results in a worse second-order approximations.

Finally, we also need an approximation for the integral
\begin{equation}
      I_3(z)
      =
    \frac{2}{\sqrt \pi \, \zeta^{\frac{5}{3}}(3/2) }
      \int_0^\infty
      \sqrt \epsilon \, d \epsilon \,
      \ln ( 1 - e^{- \omega} )
      \>.
\end{equation}
We note that the derivative of $I_3(z)$ with respect to $z$ can be written in terms of $I_1(z)$ and $I_2(z)$, as
\begin{equation}
      \frac{d I_3(z)}{d z}
      =
    \frac{2}{\sqrt \pi \, \zeta^{\frac{5}{3}}(3/2) }
      \int_0^\infty
      \frac{\epsilon}{\omega} \,
      \frac{\sqrt{\epsilon} \ d \epsilon}{e^\omega - 1}
      \propto
      I_1(z) - I_2(z)
      \>.
\end{equation}
Therefore, we can write
\begin{equation}
      I_3(z)
      =
    \frac{1}{\zeta^{\frac{2}{3}}(3/2)} \,
    \Bigl \{
      b_0'
      +
      \int_0^z d z' \,
      \bigl [ I_1(z') - I_2(z') \bigr ]
    \Bigr \}
      \>,
\end{equation}
where
$
      b_0'
      =
      - \zeta(5/2) / \zeta(3/2)
$.
Thus, we obtain the second-order approximation of $I_3(z)$ as
\begin{equation}
      I_3(z)
      \approx
    \frac{1}{\zeta^{\frac{2}{3}}(3/2)} \,
    \Bigl (
      b_0'
      + z
      - \frac{4b_1}{3} z^{3/2}
      + \frac{3b_1}{4} z^2
      \Bigr )
      \>.
\end{equation}
The first- and second-order approximations of $I_3(z)$ are illustrated in Fig.~\ref{Fig_I3}.

%
\bibliography{johns}

\end{document}